# Cooperative System of Emission Source Localization Based on SDF

Jan M. Kelner
*Institute of Telecommunications – Faculty of Electronics*
*Military University of Technology*
Warsaw, Poland
jan.kelner@wat.edu.pl

*Abstract*—Efficient and precise location of emission sources in an urbanized environment is very important in electronic warfare. Therefore, unmanned aerial vehicles (UAVs) are increasingly used for such tasks. In this paper, we present the cooperation of several UAVs creating a wireless sensor network (WSN) that locates the emission source. In the proposed WSN, the location is based on spectrum sensing and the signal Doppler frequency method. The paper presents the concept of the system. Simulation studies are used to assess the efficiency of the cooperative WSN. In this case, the location effectiveness for the WSN is compared to the single UAV.

*Keywords—cooperation system, location, localization, SDF method, signal Doppler frequency, spectrum sensing, unmanned aerial vehicle, urban environment, wireless sensor network.*

## I. Introduction

A development of modern wireless communication systems results from a dynamic development of micro- and nano-electronics, which enabled the use of higher frequency ranges. Until the mid-twentieth century, the radio communications were mainly the domain of civil services, army, aviation, and navy. In civilian applications, the wireless communication was mainly used in broadcasting systems for radio and television, and amateur radio communication. At that period, the very- (VHF) and ultra-high frequency (UHF) bands were mainly used. Currently, the rapid development of microwave technology has allowed the use of the higher frequency ranges, including for the needs of commercial cellular systems and wireless networks operating in unlicensed bands. This approach increases the density of emission sources, especially in urbanized areas. This is also related to the smaller transmitter range, which is partly due to higher attenuation of the microwave, millimeter, and terahertz waves than for VHF and UHF waves.

In the urban areas, carrying out radio reconnaissance for the needs of electronic warfare is very difficult. This results from nature of a propagation environment and the modern wireless systems. On the other hand, recent armed conflicts show that combat operations in this environment are becoming more and more common. For this reason, the development of new effective methods for the electronic warfare in urban areas is extremely important.

Spectrum monitoring and the location of emission sources are the main tasks carried out as part of the radio reconnaissance. Until recently, these tasks were performed only by specialized units that had dedicated equipment. At present, most of the units and even single soldiers are often equipped with modern electronic devices that enable the implementation of selected aspects of the electronic warfare. The ubiquitous use of the wireless communication systems, Internet of Things and unmanned platforms is conducive to this trend.

The spectrum monitoring in a limited range will be implemented by most modern transceivers based on software-defined (SDR) and cognitive radio (CR) [1–5]. The use of these technologies allows for the spectrum sensing in operating frequency of the receivers. In addition, each node of the CR network performs spectrum sensing to search for free spectral resources, evaluate currently used, and available backup channels [6,7].

The location assessment of emission sources for the needs of the electronic warfare is carried out by highly specialized radio direction-finders (DF) and localization stations. The main disadvantages of the DFs are high price and low accuracy of location, especially in urbanized areas. Primarily, the DFs are based on analysis of angle of arrival (AOA) of received signals and allowed only to determine a direction to the emission source [8,9]. To locate this source, the use of a minimum of two DFs is necessary. In this case, their mutual space orientation to the localized object significantly affects a location error for the AOA method. It should be highlight that the accuracy of radio bearing is determined in an angular measure. Therefore, the location error expressed in a linear measure increases with the distance between the source and the DF. Additionally, the use of bearing methods in the urban environment has a negative influence on their accuracy, which results from multipath propagation [10]. Some localization stations allow locating the objects using time methods, i.e., time of arrival or time difference of arrival [9,11]. In this case, several stations are also necessary for carrying out the location process. However, the accuracy of the time methods is closely related to a bandwidth of the signal emitted by the localized source. On the other hand, these location systems require synchronization, aggregation and joint processing of recorded data. This significantly influences on their complexity level and price.

The localization systems are increasingly implemented on manned and unmanned aircrafts. In this case, an ease of moving the system relative to the located object gives the possibility of faster and more accurate localization. A survey of location techniques with the mobile receiver is shown in [12].

One of the methods indicated in this review is the signal Doppler frequency (SDF) [13]. This method based on the Doppler effect enables the location of the emission sources by a single locating receiver placed on a mobile platform. Heretofore, SDF location has always been carried out on using the single platform. However, SDF use in the urbanized environment decreases the accuracy of this method [14] just as other location techniques. The purpose of this paper is to present an innovative application of SDF in a cooperative system of locating the emission sources in the urban areas. Elements of this system are arranged on several unmanned aerial vehicles (UAV) and operated as a wireless sensor network (WSN) [15,16]. This solution provides an increase of monitoring area, accuracy and speed of location process.

The remainder of this paper is organized as follows. A problem formulation based on a spatial scenario is shown in Section II. Section III presents a description of the cooperative system, the classical SDF method, and modification of a location algorithm. The effectiveness assessment of the novel solution is made on the basis of simulation studies. Their exemplary results are shown in Section IV. Section V contains the summary of the paper.

## II. PROBLEM FORMULATION

The location of the emission sources in the urbanized environment is difficult due to the multipath propagation of radio waves. This phenomenon is the reason for the fading and dispersion of the received signals in time, frequency, and angle domains [17,18]. In addition, in these areas, high density of wireless cellular networks, WiFi, and other radio transmitters may cause additional interferences. Thus, carrying out the locating procedures by the single mobile platform is usually difficult and inefficient.

Figure 1 shows the spatial scenario of an analyzed problem. In this case, we assume that the localized emission source is located on a wheeled vehicle that can move in the urban area. Its transmitter works on a specific carrier frequency. The WSN consisting of the co-operating UAVs is aimed to locate this emission source quickly and accurately. Each UAV is equipped with the SDR receiver that works in the operating band of the localized transmitter. The algorithms of spectrum sensing and SDF localization are implemented in each receiver.

## III. PROPOSAL TO SOLVE PROBLEM

In this Section, we present a description of the concept of UAV cooperation within the WSN used to locate the emission source. Then, the short characteristics of the classic SDF method and the location algorithm modification are shown.

### A. Cooperative Location System Based on UAV WSN

The cooperative location system consists of several UAVs forming the WSN. The system performs the tasks set by the unit command consisting in locating the emission source in the urban area. Each UAV is equipped with the SDR receiver, in which SDF location algorithm is implemented. A navigation system should be an additional UAV equipment.

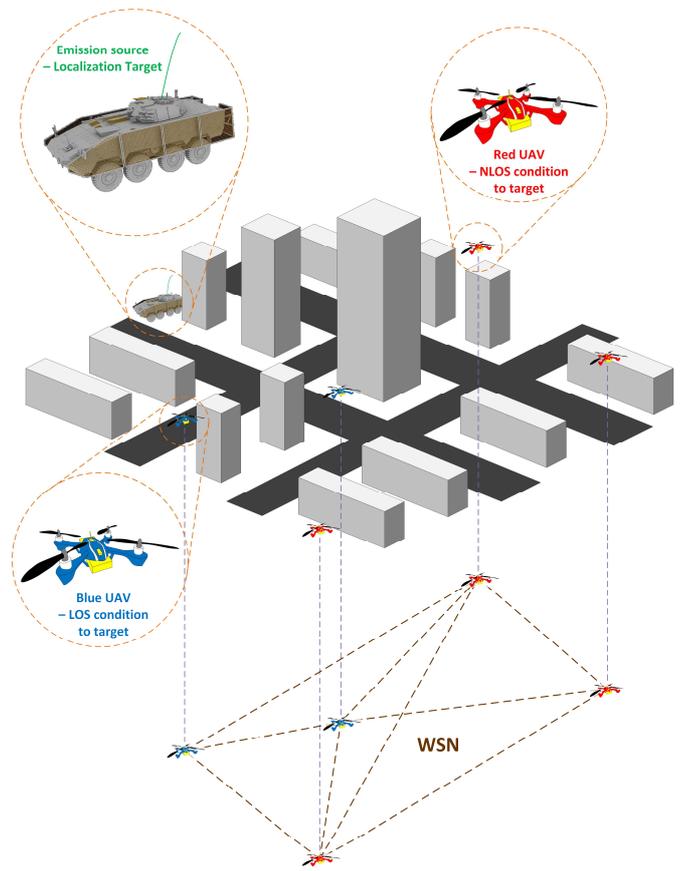

Fig. 1. Emission source localization by cooperative UAVs operating as WSN – spatial scenario.

As the UAV navigation system, receiver of a global navigation satellite system (GNSS) integrated into an inertial navigation system (INS) is usually used [19,20]. However, the use of these systems in a war zone may be difficult due to jamming [21] or spoofing [22] GNSS signals. For this purpose, anti-jamming systems [23,24] and anti-spoofing [25] may be used. Another approach is to equip the UAV with a navigation receiver for a dedicated short-range terrestrial system. The concept of such a system based on SDF is presented in [26,27].

In an ideal solution, we can assume that UAVs are autonomous and can automatically locate emission sources. In this case, the UAV should be additionally equipped with a visual-based navigation system (VBNS) [28], e.g., based on simultaneous localization and mapping (SLAM) algorithms [29,30]. In this configuration, the radio navigation system is used to determine the positions of the UAV and localized emission source in geographical coordinates, as well as to maintain the direction of searching the urban area. Whereas, the VBNS is primarily responsible for avoiding buildings and others obstacles, and maneuvering in the urban environment.

Additionally, the standard equipment of each UAV is a transceiver, which provides communication with the unit command and other UAVs cooperating within the WSN. Presently, most commonly used SDR transceivers have several receiving inputs and transmitting outputs. Hence, such a transceiver can also operate as the SDR receiver used for spectrum sensing and localization, as well as transmitter to communicate with the command and other WSN nodes. It is also possible to adapt this transceiver to the dedicated navigation system. For a limited number of inputs/outputs, the SDR transceiver can operate in either time or frequency division multiple access modes. A video camera is used in a potential VBNS and may transmit vision to the command from a particular area of operation. So, UAV can carry out additional image recognition.

Figure 2 presents a tactical scenario for a hypothetical city. The blue-and-red dotted line shows the line of contact of armies. The target of the cooperative WSN is to determine the location of the emission source operating in a specific frequency channel and marked by a red dot. The WSN nodes, i.e., UAVs, start from the place marked by a blue dot in a distributed configuration. In the beginning, another direction of searching the area is designated for each UAV. This direction can be modified depending on the obstacles encountered in the environment or as a result of the location algorithm.

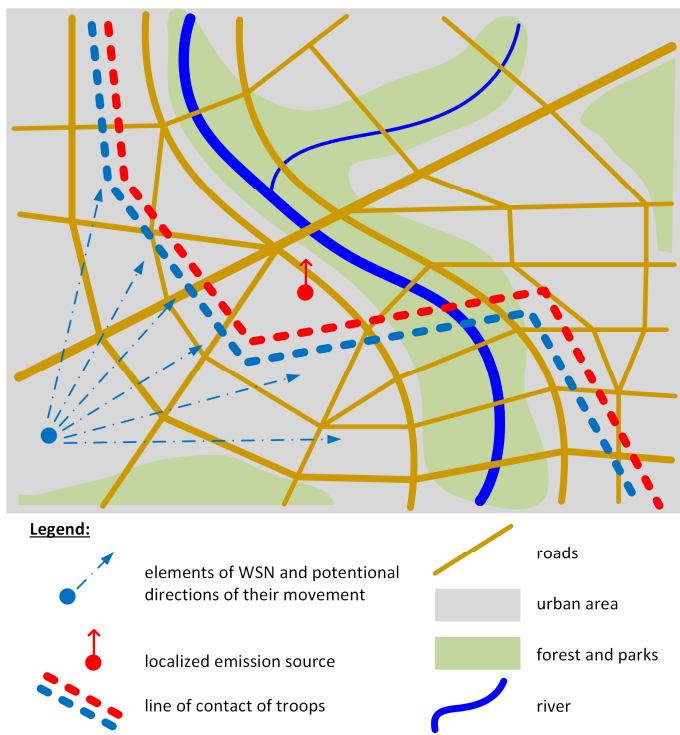

Fig. 2. Tactical scenario for the hypothetical city.

### B. Classic SDF Method

The classic approach to SDF was repeatedly presented in literature, e.g., in [13,31]. An overview of SDF research is presented in [32], while [33] provides a survey of SDF applications in location and navigation.

SDF is based on the analytical solution of a wave equation for the moving emission source [34]. We assume that the transmitter located at the point $(x_0, y_0, z_0)$ emits a signal on the carrier frequency $f_0$. The receiver moving at speed, v, may locate the transmitter based on the Doppler frequency shift (DFS) in the received signal, $f_D(t)$. Then, the coordinates of the signal source position are estimated based on the following formulas [13,31–34]:

$$\begin{cases} \tilde{x} = v \dfrac{t_1 A(t_1) - t_2 A(t_2)}{A(t_1) - A(t_2)} \\ \tilde{y} = \pm \sqrt{\left[ v \dfrac{(t_1 - t_2) A(t_1) A(t_2)}{A(t_1) - A(t_2)} \right]^2 - \tilde{z}^2} \end{cases} \quad (1)$$

where $A(t) = \sqrt{1 - F^2(t)} / F(t)$, $F(t) \cong f_D(t)/f_{D\max}$, $f_{D\max} = f_0 v/c$ is the maximum DFS, and c is the light speed.

The presented above version of SDF refers to a location on a plane (2D). In this case, we assume that $\tilde{z} = z_0$ is known, what corresponds, e.g., to a flight altitude of the UAV. A spatial (3D) SDF version and a way of avoiding ambiguity in (1) are presented in [32].

### C. Modification of Location Algorithm

The algorithm modification results from locating the signal source by the WSN in the urbanized environment. In the initial search period, we assume that all UAVs are moving at a considerable distance from the transmitter. Then, there are usually occurred non-line-of-sight (NLOS) conditions, and according to Fig. 1, all UAVs are red. At a certain distance from the signal source, line-of-sight (LOS) or obscured LOS (OLOS) conditions may occur for some UAVs. Considering the designations in Fig. 1, these UAVs are blue. The difference in the location method for different propagation conditions is related to the estimation of the instantaneous DFS, $f_D(t)$. In the LOS conditions, this is the classical approach, while in the NLOS conditions the analysis of the dispersive Doppler spectrum is used [14]. Detection of the used spectral analysis method depends on the received signal power in the channel used by the localized transmitter. For this aim, we use methods typical for the spectrum sensing, e.g., [6,7].

The evaluation of the emission source location can be carried out based on averaging all results from individual WSN nodes, i.e., from each UAVs. A more effective solution is to choose only those UAVs for which LOS/OLOS conditions exist. For the urban area, a similar approach is also used in other location techniques, e.g., [35,36].

## IV. SIMULATION RESULTS

The effectiveness assessment of the cooperative system is based on simulation studies carried out in the Matlab. Simulations are based on a scenario similar to that described in [14]. For NLOS conditions, we used the channel model presented in [37]. This model considers the majority of significant propagation phenomena, including, i.a., path loss, noise, fading, Doppler effect, dispersion in time, frequency, and angle domains.

In simulation studies, we adopt variability of LOS/NLOS conditions for the individual UAVs that is depicted in Fig. 3.

Depending on the propagation conditions, the statistical properties of the received signal are modeled in a different way.

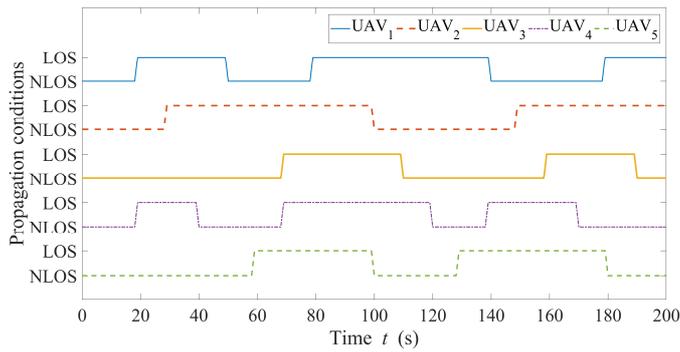

Fig. 3. Variability of LOS/NLOS conditions for individual UAVs.

Based on the spectral analysis of the received signal, DFSs are estimated. Exemplary Doppler curves, i.e., DFSs versus time, are shown in Fig. 4. In this case, we can observe larger DFS estimation errors for time intervals at which the NLOS conditions occur for the analyzed UAV.

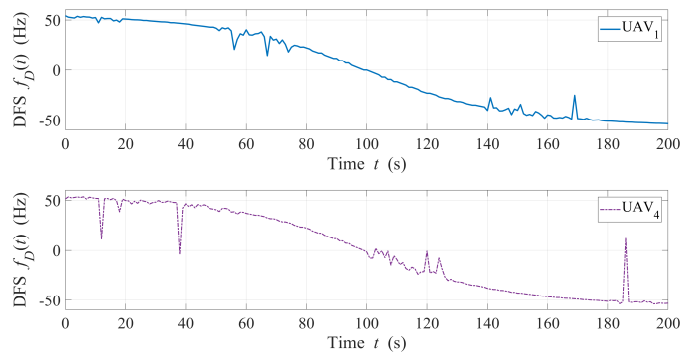

Fig. 4. Doppler curves for 1st (up) and 4th (down) UAVs.

The signal with duration 1s is used to estimate the DFS. The current position of the emission source is determined based on the last 20 DFSs. As simulation results, exemplary rms location errors for the individual UAVs are determined. Graphs of these errors are shown in Fig. 5.

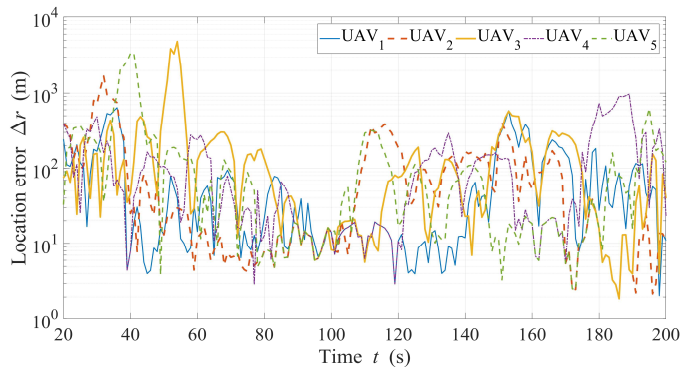

Fig. 5. Location errors versus time for single UAVs.

For the individual UAVs, the rms errors fluctuate from a few to several hundred meters. For the entire analyzed trajectory of movement, the average errors are 81, 116, 207, 130, and 182 m for UAV with numbers from 1 to 5, respectively. Therefore, the accuracy of the transmitter location by the individual platforms is characterized by a significant spread. For example, the average error for $UAV_3$ is more than two and a half times higher than for $UAV_1$.

In the cooperative system, we can use the variability of propagation conditions in favor of the location accuracy. However, the use of an arithmetic mean for the source coordinates estimated by the UAVs do not give the desired effect. In Fig. 6, a blue line marked as 'averaged' shows the location error for the arithmetic mean. In this case, the average error for the entire analyzed route is 108 m, i.e., like for the single platforms. Additionally, the minimum and maximum errors for the UAVs from Fig. 5 are shown in Fig. 6.

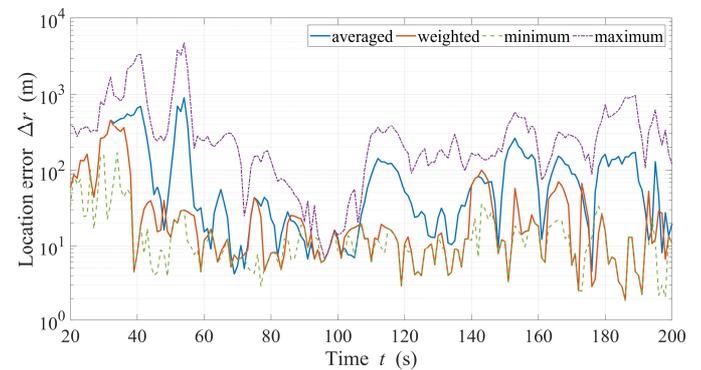

Fig. 6. Location errors in cooperation system.

The use of a weighted mean is by far a better approach, as shown by a red line in Fig. 6. For the cooperative WSN, we assumed that the UAVs share information about the received signal power and the current estimated coordinates of the located source. Based on the power comparison received by the individual platforms in the analyzed frequency channel, the propagation conditions occur on paths between each UAV and the transmitter are determined (see Fig. 3). Then, the source position is determined as the average of the coordinates estimated by the UAVs for which the LOS/OLOS conditions exist. If the NLOS conditions exist for all UAVs, then we use the typical arithmetic mean. For the weighted mean and entire analyzed route, the average error is only 38 m. This approach is improved the location accuracy for the cooperative WSN.

## V. Conclusion

In this paper, a novel way of locating the emission sources by the cooperative WSN that is based on SDF, spectrum sensing, and using the UAVs was presented. The concept of the cooperative system and the method of determining the position were shown. Simulation studies were used to evaluate the effectiveness of the proposed system. For cooperative approach, the obtained results showed that the location error could be reduced from two to five times in relation to the single platform. The simulation results showed that the source position estimation based on the proposed weighted average is characterized by a smaller error than the use of a typical arithmetic mean for all UAVs. The presented simulation results relate to a simple variant in which the focus was only on UAVs cooperation and the change of propagation conditions (LOS/NLOS). In the general case, frequency stability of the received signal or the accuracy of determining the speed, direction of movement, and position of each UAV using, e.g., GPS, should be considered. Simulation tests should also consider conditions typical of a battlefield, which force, e.g., a frequent change in the UAV motion direction or using a frequency hopping by military transceivers. Some of these aspects are presented in other papers of the author, e.g., the influence of signal frequency stability of the localized source on the positioning accuracy by the SDF method [38]. Additionally, UAVs included in the system might cooperate with the unit command, but also with single soldiers. Therefore, flying platforms could provide increased situational awareness on the battlefield, not only within the scope of the image recognition from the air but above all in the radio recognition.